\documentclass[12pt]{article}
\usepackage{epsfig}

\newlength{\oline}

\usepackage{graphicx}
\begin{document}
\begin{center}
{\large \bf  Total cross-section for photon-axion conversions in
external electromagnetic field}\\

\vspace*{1cm}

{\bf D. V. Soa$^{a,}$\footnote{dvsoa@hnue.edu.vn}, H. N.Long$^{b,}
$\footnote{hnlong@iop.vast.ac.vn}, T. D.
Tham$^{c,}$\footnote{tdtham@pdu.edu.vn}\\}

\vspace*{0.5cm}

$^a$ Department of Physics, Hanoi University of Education, Hanoi, Vietnam \\
$^b$ Institute of Physics, Vietnam Academy of Science and Technology, \\
10 Dao Tan, Ba Dinh, Hanoi, Vietnam \\
$^c$  Pham Van Dong University, 986 Quang Trung Street, Quang Ngai City, Vietnam\\

\vspace{1cm}
 {\bf Abstract}\\
\end{center}
\hspace*{0.5cm} We reconsider the conversion of the photon into
axion in the external electromagnetic fields, namely in the static
fields and in a periodic field of the wave guide. The total
cross-sections for the conversion are evaluated in detail. The
result shows that with strong strength of external electromagnetic
fields, the cross-sections are large enough to measure the axion
production. In the wave guide there exists the resonant conversion
at the low energies, in which the value of cross-sections is much enhanced.\\[0.2cm]
\newpage
\section{Introduction}
The Peccei - Quinn (PQ) mechanism~\cite{pec,pec1} provides a
simple explanation for the strong CP problem by the introduction
of a light pseudoscalar particle, called the axion
~\cite{wei,wil}, which receives a small coupling to
electromagnetism (similar to neutral pion in QCD).  At present,
the axion mass is constrained by laboratory searches
~\cite{kim,cheng,pec2} and by astrophysical and cosmological
considerations~\cite{tur,raf,kol} to between $10^{-6}$ eV and
$10^{-3}$ eV. If the axion has a mass near the low limit of order
$10^{-5}$ eV, it is a good candidate for the dark matter of the
Universe. It was also argued that an axion- photon oscillation can
explain the observed dimming supernovas if it has a rather small,
order of $10^{-16}$ eV
~\cite{per}.\\
Neutral pions, gravitons, hypothetical axions, or other light
particles with a two-photon interaction can transform into photons
in external electromagnetic (EM) fields, an effect first discussed
by Primakoff ~\cite{pri}. This effect is the basis of Sikivie's
methods for the detection of axions in a resonant
cavity~\cite{sik}. The experiment CAST(Cern Axion Solar
Telescope)~\cite{ira,cast,zio} at CERN searches for axions from
the sun or other sources in the Universe. The experiment uses a
large magnet from LHC to convert solar axions into detectable
X-ray photons. The potential of the CAST experiment for exotic
particles was discussed in Refs.~\cite{ daf,fer}.\\
In our previous works~\cite{long,int}, we have calculated the
different cross-sections for the photon-axion conversion in
external EM fields in detail. However, some numerical evaluations
in comparison with the experiments are not realistic. The purpose
of this paper is to evaluate {\it the total cross sections} for
the photon-axion conversion in external EM fields, including the
static fields  as in Ref.~\cite{long} and also the periodic
 field of the wave guide~\cite{int}.\\
\hspace*{0.5cm}Consider the conversion of the photon $\gamma$ with
momentum $q$ into the axion $a$ with momentum $p$ in an external
electromagnetic field. The matrix element is given
by~\cite{long,int}
\begin{equation}
 \langle p|{\cal
M}|q\rangle = -\frac{g_{a\gamma}}{2(2\pi)^2
\sqrt{q_0p_0}}\varepsilon_{\mu}(\vec{q},\sigma)\varepsilon^{\mu\nu\alpha\beta}q_{\nu}
\int_{V}
e^{i\vec{k}\vec{r}}F_{\alpha\beta}^{class}d\vec{r},\label{equa1}
\end{equation}
where $ \vec{k}\equiv \vec{q}-\vec{p}$ is  the momentum transfer
to the EM field, $\varepsilon^{\mu}(\vec{q},\sigma)$ represents
the polarization vector of the photon, and $g_{a\gamma}\equiv
g_{\gamma}\frac{\alpha}{\pi f_a}$= $g_{\gamma}\alpha
m_a(m_u+m_d)(\pi f_{\pi}m_{\pi}\sqrt{m_um_d})^{-1}$. In particular
in the Dine- Fischler - Srednicki - Zhitnitskii model
~\cite{din,zhi}: $g_{\gamma}$(DFSZ)$\simeq 0.36$, and in the Kim-
Shifman- Vainshtein- Zakharov model ~\cite{kim3,shi} (where the
axions do not couple to light quarks and  leptons):
$g_{\gamma}$(KSVZ)$\simeq -0.97$.\\
\section{Conversions in the electric field}
 We reconsider the photon-axion conversion in  the homogeneous
electric field of the flat condenser of size $l_x\times l_y\times
l_z$ ( instead of $a\times b \times c$ in Ref.~\cite{long}) .
Using the coordinate system with the $x$ - axis parallel to the
direction of the field, i.e., $F^{10}=-F^{01} = E$, where $E$ is
the strength of electric field. From Eq.(\ref{equa1}) we obtained
the differential cross section for the conversion~\cite{long}
\begin{equation}
\frac{d\sigma^e(\gamma \rightarrow a)}
{d\Omega}=\frac{g^2_{a\gamma}E^2}{2(2\pi)^2}\left[\frac{\sin(\frac{1}{2}
l_x k_{x})\sin(\frac{1}{2}l_y k_{y})\sin(\frac{1}{2} l_z
k_{z})}{k_{x}k_{y}k_{z}}\right]^{2}\left(q_{y}^{2}+q_{z}^{2}\right).\label{equa2}
\end{equation}
From Eq.(\ref{equa2}) we see that if the photon moves in the
direction of the electric field i.e., $q^{\mu} = (q, q, 0,0)$ then
the different cross-section vanishes. If the momentum of the
photon is parallel to the $y$ - axis (the orthogonal direction),
i.e., $q^{\mu} =( q, 0, q, 0 )$ then Eq.(\ref{equa2}) becomes
\begin{eqnarray}
\frac{d\sigma^e(\gamma \rightarrow a)} {d\Omega'}&=&\frac{32
g^2_{a\gamma}E^2q^2}{(2\pi)^2}[\sin(\frac{l_x}{2} psin\theta\
\sin\varphi')\sin(\frac{l_y}{2}(q-p\cos\theta)) \nonumber \\
&\times&\sin(\frac{l_z}{2}
p\sin\theta\cos\varphi')]^{2}(p^{2}\sin^{2}\theta
\sin\varphi'\cos\varphi'(q-p\cos\theta))^{-2},\label{equa3}
\end{eqnarray}
where $\varphi'$ is the angle between the $z$ - axis and the
projection of $\vec{p}$ on the $xz$ - plane.
 For the forward scattering case ($\theta \approx 0$) we have
\begin{equation}
\frac{d\sigma^e(\gamma \rightarrow a)}{d\Omega'}=\frac{2
g^2_{a\gamma}E^2l_x^2l_z^2}{(2\pi)^2\left(1-\sqrt{1-\frac{m^2_a}{q^2}}\right)^2}
\sin^2\left[\frac{q
l_y}{2}\left(1-\sqrt{1-\frac{m^2_a}{q^2}}\right)\right].\label{equa4}
\end{equation}
 In the limit $m_a^2 \ll q^2$ and  $l_y\sim m^{-1}_a$ , from Eq.(\ref{equa4}) we have
\begin{equation}
\frac{d\sigma^e(\gamma \rightarrow a)}{d\Omega'}\simeq\frac{
g^2_{a\gamma}E^2l_x^2l_z^2}{16\pi^2}. \label{equa5}
\end{equation}
 We can see from Eq.(\ref{equa5}) that in this case {\it the
cross section does not depend on the provided photon energies}.
 It is noticed that in our previous work~\cite{long} we have obtained
Eq.(\ref{equa3}) and evaluated the different cross sections in
detail, however some numerical evaluations in comparison with
the experiments are not realistic. \\
Now we are mainly interested in {\it the total cross-section}
$\sigma^e(q)=\int d \Omega (d \sigma^e /d\Omega)$ from the general
formula (\ref{equa3}). For this purpose, we  note that the
integrand as well as the total cross-section depend on provided
photon momentum $q$ (at least larger than the axion mass) are very
rapidly oscillated with $q$. To overcome difficulties, we plot a
large spectrum of points $(q,\sigma^e(q))$ corresponding to a
large number of values of $q$ in the interested domain. The
orientation of the spectrum will reflect the correct variation of
the cross-section. The parameters are chosen as
follows~\cite{long}, $l_x =l_y =l_z =1m$, the intensity of the
electric field $E = \frac{100 kV}{m}$ and the axion mass $m_a =
10^{-5} eV$ (nearby the low limit of mass window~\cite{tur}). The
total cross-section on the selected range of the provided momenta,
$q=10^{-4}\div 10^{-3}$ eV, are given in Figure~\ref{tab1}. The
upper plot is depicted as 300 points, and the lower one is for
3000 points. As demonstrated in the two plots, when the number of
points is increased then  the resonances become shapely, in which
the cross-sections are quite large ($\sigma \sim 10^{-29} cm^2$).
 The numerical evaluation for Eq.(\ref{equa5}), the cross-section
is given by $\frac{d\sigma^e(\gamma
\rightarrow a)}{d\Omega'}\simeq 1.8\times 10^{-37} cm^2$.\\
It is noticed that this parameter is the derivative one, not
concerning as any characteristic scales of the model. Its value
depends only on a choice of the parameters such as the axion mass,
the size of condenser, the field strength, and so on. Let us
remark that when the momentum of photon is perpendicular to the
electric field $E$ we have then the most optimal condition for the
experiments.
\section{Conversions in the magnetic field }
\hspace*{0.5cm} We move on to conversions in the strong magnetic
field of the solenoid with a radius $R$ and a length $h$.
 If the momentum of the photon is parallel to the $x$ - axis,
 the different cross-section is given by~\cite{long}
\begin{eqnarray}
\frac{d\sigma^m(\gamma\rightarrow a)}{d\Omega'}&=&2g^2_{a\gamma}
R^2B^2J_1^2\left(Rq\sqrt{\left(1-\cos\theta\sqrt{1 -
\frac{m_a^2}{q^2}}\right)^2 +
\left(1-\frac{m_a^2}{q^2}\right)\sin^2\theta\cos^2\varphi'}\right)\nonumber\\
&&\times\left[\left(1-\cos\theta\sqrt{1-\frac{m_a^2}{q^2}}\right)^2
+ \left(1-\frac{m_a^2}{q^2}\right)\sin^2\theta\cos^2\varphi '
\right]^{-1}q^{-2}  \nonumber \\
& &\times\sin^2\left(\frac{hq}{2}\sqrt{1 -
\frac{m_a^2}{q^2}}\sin\theta\sin\varphi'\right)\left[(1-\frac{m_a^2}{q^2})
\sin^2\theta\sin^2\varphi'^2\right]^{-1}, \label{equa7}
\end{eqnarray}
where $B$ is the strength of magnetic field and $J_1$ is the
spherical Bessel function of the first kind. For the forward
scattering case and in the limit $m_a^2 \ll q^2$,
 $R\leq m^{-1}_a$, we have
\begin{equation}
\frac{d\sigma^m(\gamma\rightarrow a)}
{d\Omega'}\simeq\frac{1}{2\pi^2}g^2_{a\gamma}V h B^2,\label{equa8}
\end{equation}
where $V$ is the volume of the solenoid. From (\ref{equa8}) we see
that {\it the conversion probability is proportional to the square
of the field strength, the active length and the volume of the solenoid}.\\
 To evaluate the total cross-section from the
general formula (\ref{equa7}), the parameter values are given as
before and the remaining ones are chosen as follows: $R=l=1\
\mathrm{m}=5.07\times 10^6\ \mathrm{eV}^{-1}$ and $B = 9\
\mathrm{Tesla}=9\times 195.35\ \mathrm{eV}^2$ ~\cite{cast}.
 The total cross-sections on the selected range of
momenta $q$ is presented in Figure \ref{tab2}. The upper plot is
depicted as 300 points, and the lower one is for 3000 points. From
the figures we see that the total cross-sections for the axion
production in the strong magnetic field ($\sigma \sim 10^{-19}
cm^2$ ) are much bigger than that in the electric field. This
happens because of $B\gg E$. Numerical evaluation for the formula
(\ref{equa8}), the cross-section is given by
$\frac{d\sigma^m(\gamma \rightarrow
a)}{d\Omega'}\simeq 1.1\times 10^{-32} cm^2$.\\
\section{Conversions in the  wave guide}
 \hspace*{0.5cm} Now  we are interested in  conversions in the external EM
field of the $TE_{10}$ mode of the wave guide with frequency equal to the axion
mass ~\cite{int}. The nontrivial solution of the $TE_{10}$ mode is given by \\
\begin{eqnarray}
H_z &=& H_o\cos\left(\frac{\pi x}{l_x}\right)e^{ikz-i\omega t},\nonumber \\
H_x &=& -\frac{ikl_x}{\pi}H_o\sin\left(\frac{\pi x}{l_x}\right)
e^{ikz-i\omega t},\nonumber \\
E_y &=& i\frac{\omega a \mu}{\pi}H_o\sin\left(\frac{\pi
x}{l_x}\right) e^{ikz-i\omega t}.\label{equa9}
\end{eqnarray}
Here the propagation of the EM wave is in the $z$ - axis. If the
momentum of the photon is parallel to the $x$ - axis, then
 the different cross-section is given by
 the $x$ - axis, then
\begin{eqnarray}
\frac{d\sigma(\gamma \rightarrow a)} {d\Omega'}&=&\frac{8
g^2_{a\gamma}H_0^2l^2_x q^2}{\pi^4}( 1 +
\frac{\omega}{q})\left[\omega (q -p\cos\theta) -
\frac{\pi^2}{l^2_x }\right]^2
\nonumber \\
& &\times\left[\frac{\cos\frac{l_x}{2}(q-p\cos\theta)
\sin\frac{l_y}{2}(p\sin\theta\cos\varphi')\sin\frac{l_z}{2} (-
p\sin \theta\sin\varphi'+ k
)}{[(q-p\cos\theta)^2-\frac{\pi^2}{l^2_x}] .
p\sin\theta\cos\varphi'( - p\sin \theta\sin\varphi'+ k
)}\right]^2. \label{equa11}
\end{eqnarray}
To evaluate the total cross-section for the formula (\ref{equa11})
, we take $H_0 = B$, $\omega = m_a =10^{-5}\ \mathrm{eV}$. The
remaining parameters are chosen as before. Figure 3 shows the
dependence of the total cross -section $\sigma$ as a function of
momentum $q$. The moment range is chosen at the lower values,
$q=10^{-5}\div 10^{-4}$ eV. We can see from the figure that there
exists {\it a main resonant conversion } at the value $q =
5.1\times 10^{-5}\ \mathrm{eV}$, the cross section is given by
$\sigma \simeq 10^{-17}cm^2$, which is much bigger than those in
the static fields. {\it This is the best case for photon - axion
conversions}. We note that the inverse process of the axion-photon
conversion ($a\rightarrow\gamma $) is also important for the axion
detection in experiments, which was calculated in detail in Ref.\cite{pie}.\\
 \section{Conclusion}

In this paper  we have reconsidered the conversion of the photon
into axion in the external electromagnetic fields, namely in the
static fields and in a periodic field of the wave guide. The
numerical evaluations of the total cross-section are also given in
detail. Our result shows that with the strong strength of external
electromagnetic fields, the cross-sections are large enough to
measure the axion production. In the wave guide there exists the
resonant conversion at the low energies, in which the value of cross-sections is much enhanced. \\
\hspace*{0.5cm}Finally, in this work we have considered only a
theoretical basis for the experiments, other techniques concerning
construction and particle detection can be found in
Refs~\cite{cast,zio,klau}.\\

\hspace*{1cm}
 \textbf{Acknowledgements:} This research  is
funded in part by Vietnam National Foundation for Science and
Technology Development(NAFOSTED) under grant No. $103.03-2012.80$.\\

\newpage

\begin{figure}
  \includegraphics[width=12cm,height=7cm]{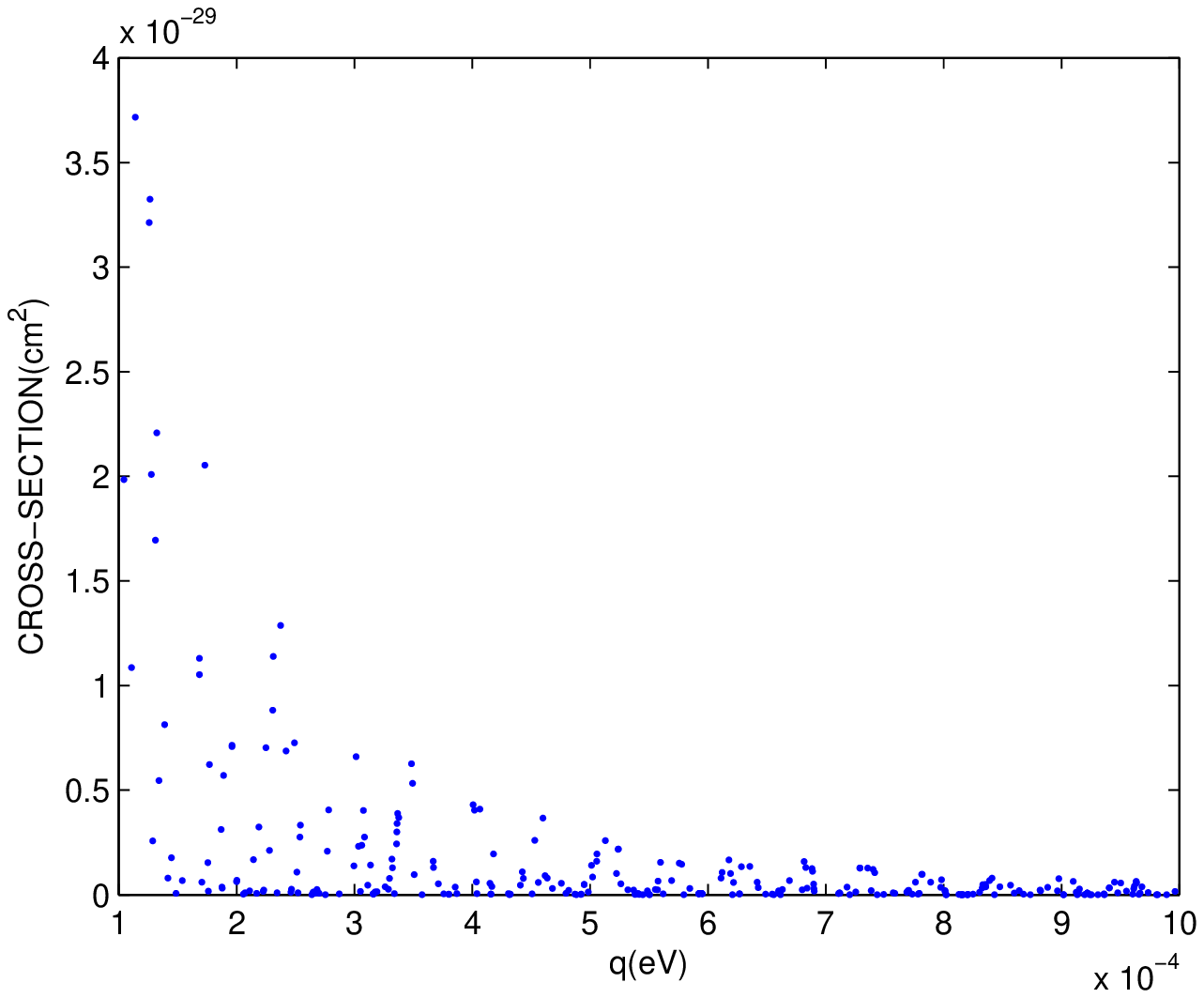}\\
  \includegraphics[width=12cm,height=7cm]{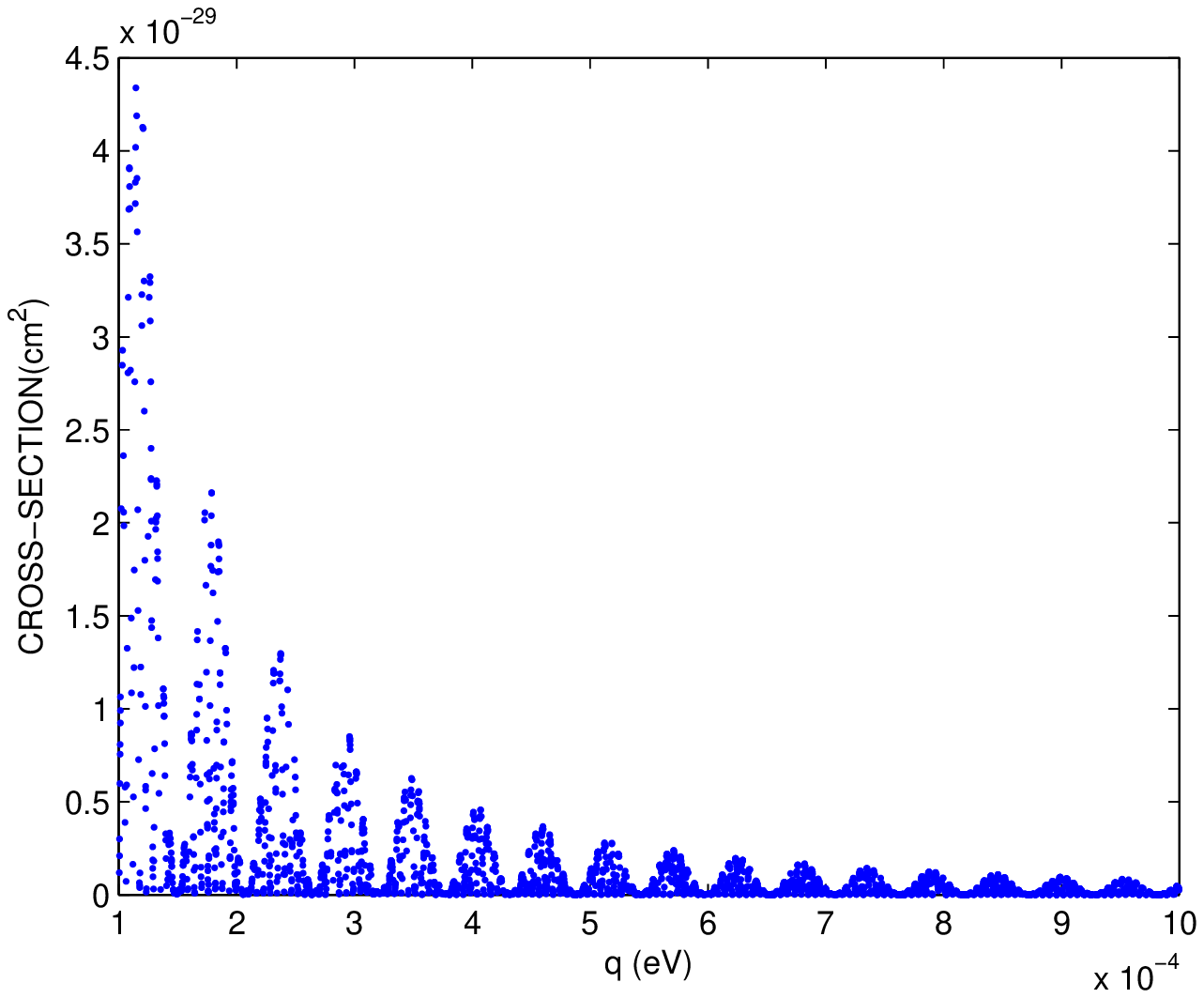}
  \caption{The total cross-section for the photon-axion
conversion in an electric field as a function of provided
momentum, $q=10^{-4}\div 10^{-3}$ eV. The upper plot is depicted
as 300 points, and the lower one is for 3000 points.}\label{tab1}
\end{figure}

\begin{figure}
  \includegraphics[width=12cm,height=7cm]{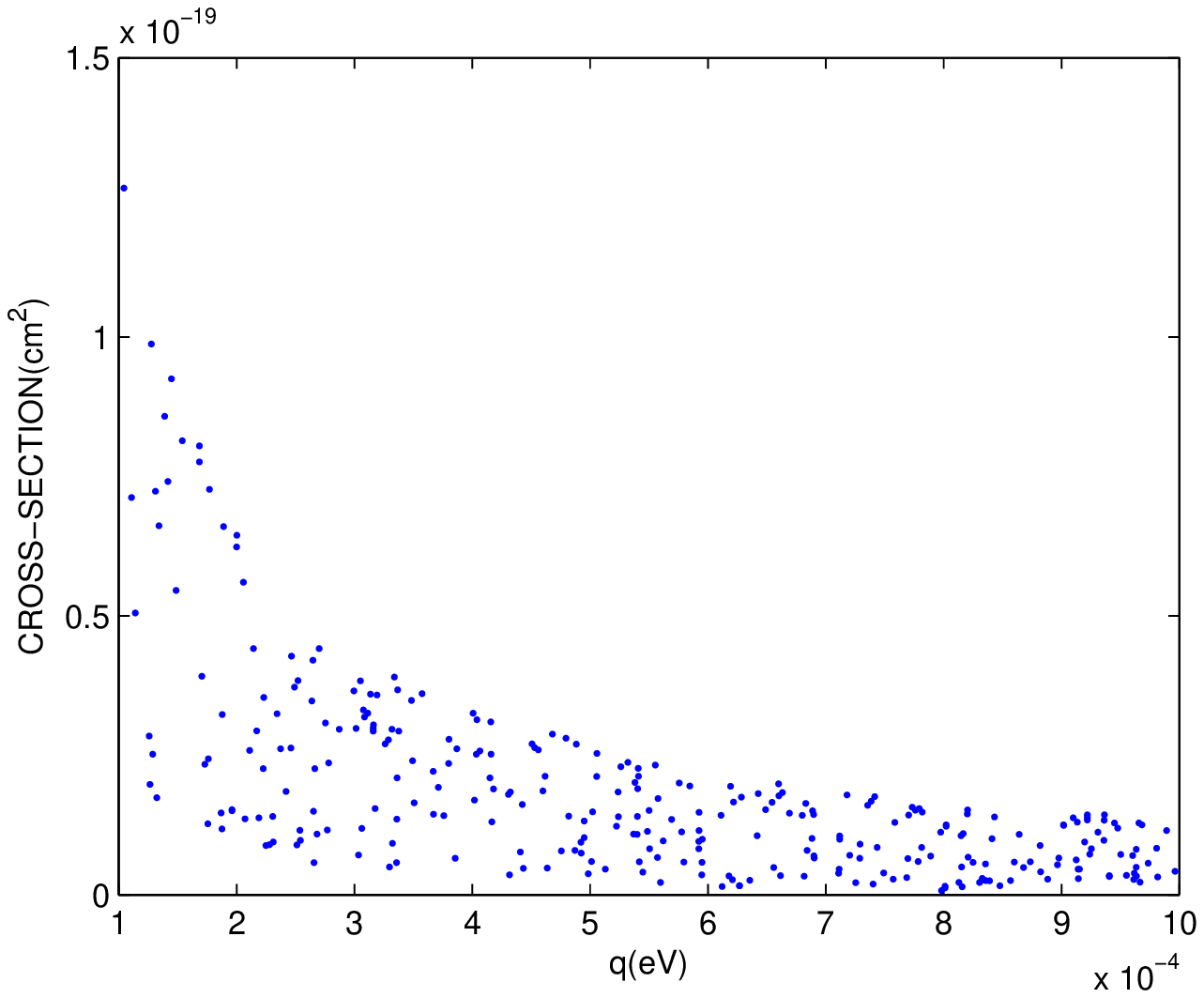}\\
  \includegraphics[width=12cm,height=7cm]{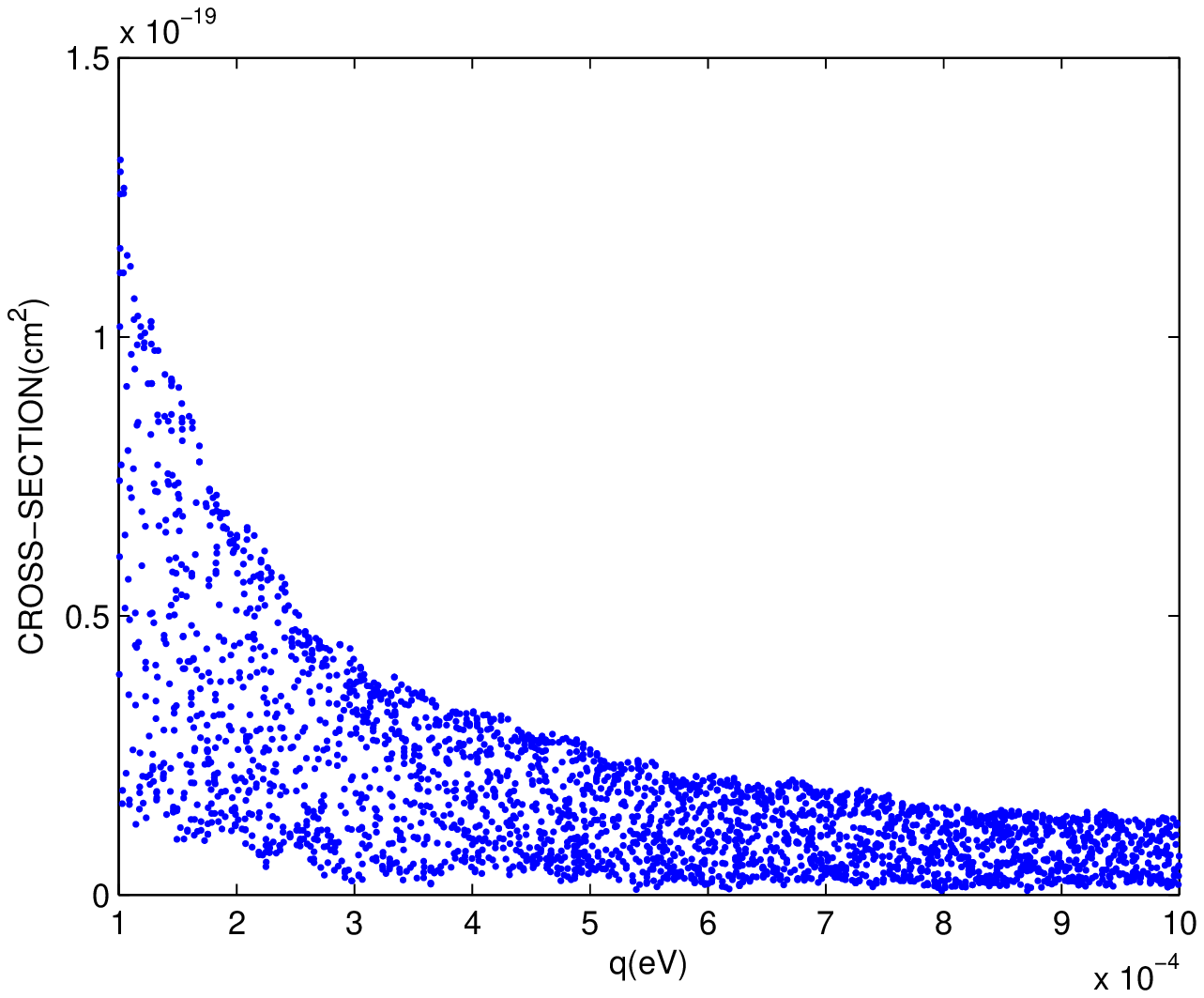}
  \caption{The total cross-section for the photon-axion
conversion in the magnetic field as a function of provided
momentum. The upper plot is depicted as 300 points, and the lower
one is for 3000 points.}\label{tab1}
\end{figure}

\begin{figure}

  \includegraphics[width=12cm,height=7cm]{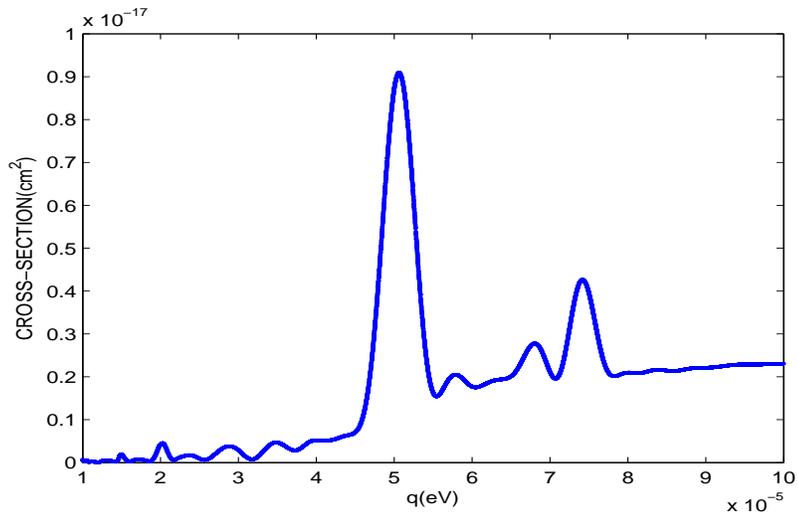}
  \caption{The total cross-section for the photon-axion
conversion in the wave guide as a function of provided momentum.
The moment range is chosen at the low values, $q=10^{-5}\div
10^{-4}$ eV. }\label{tab2}
\end{figure}

\end{document}